\documentclass{PoS}
\usepackage{natbib} 
\usepackage{graphicx}
\usepackage{txfonts}
\def\jA{IGR~J08262-3736}
\def\jB{IGR~J17348-2045}
\def\jC{SAX\,J1818.6-1703}
\def\jD{IGR~J17354-3255}
\def\jE{IGR~J16328-4726} 
\def\inte{{\em INTEGRAL}}
\def\xmm{{\em XMM-Newton}}

\def\chan{{\em Chandra}}

\def\beppo{{\em BeppoSAX}}

\def\swift{{\em Swift}}

\def\ferg{erg~cm$^{-2}$~s$^{-1}$}

 \usepackage{multicol}

\title{Soft X-ray follow-up of five hard X-ray emitters}

\ShortTitle{Follow-up of 5 hard-X emitters}

\author{\speaker{L.Pavan}$^{1}$, E. Bozzo$^{1}$, C. Ferrigno$^{1}$, M. Falanga$^{2}$,
  S. Campana$^{3}$, S. Paltani$^{1}$, L. Stella$^{4}$, 
  R. Walter$^{1}$ \\
       $^{1}$ ISDC - Universit\`e de Gen\`eve.\\
        Chemin d'Ecogia 16, CH-1290 Versoix, Switzerland \\
       $^{2}$  International Space Science Institute (ISSI), Switzerland\\
        $^{3}$  INAF - Osservatorio Astronomico di Brera, Italy\\
       $^{4}$  INAF - Osservatorio Astronomico di Roma, Italy\\
       E-mail: \email{Lucia.Pavan@unige.ch}}

\abstract{
We studied the soft-X-ray emission of five hard-X sources:
\jA, \jD, \jE, \jC\ and \jB. 
These sources are: a confirmed supergiant high mass X-ray binary (\jA); 
candidates (\jD, \jE) and confirmed (\jC) supergiant fast X-ray transients; 
\jB\ is one of the as-yet unidentified objects discovered with
INTEGRAL. \\
Thanks to dedicated \xmm\ observations, we obtained the first detailed
soft X-ray spectral and timing study of \jA.
The results obtained from the observations of \jD\
and \jE\ provided further support in favor of their association with
the class of Supergiant Fast X-ray Transients. 
\jC, observed close to phase 0.5, was not detected by \xmm, thus supporting the idea that this source 
reaches its lowest X-ray luminosity ($\sim10^{32}$ erg/s) around apastron. 
For \jB\, we identified for the first time the soft X-ray counterpart
and proposed the association with a close-by radio object, suggestive
of an extragalactic origin.
In this proceeding we discuss the results obtained from the \xmm\ follow-up
observations of all the five sources.}

\FullConference{"An INTEGRAL view of the high-energy sky (the first 10 years)"
9th INTEGRAL Workshop and celebration of the 10th anniversary of the launch,\\
		October 15-19, 2012\\
		Bibliotheque Nationale de France, Paris, France}

\begin{document}

\section{Introduction}

High-mass X-ray binaries (HMXBs) comprise a compact object orbiting a
massive O-B spectral type star. The compact object,  
usually a neutron star (NS), emits a conspicuous amount of X-ray radiation 
(up to luminosities of $\sim$10$^{37}$~erg/s) due to the accretion of
matter from the OB companion.  
Depending on the nature of the giant star, HMXBs are classified 
as Be (BeXBs) or supergiant (SGXBs) X-ray binaries.
While in the first case the NS is orbiting around the companion in a
highly eccentric orbit \citep[see e.g.][]{stella86};
in SGXBs the compact object moves around its companion 
in a nearly circular orbit.
The luminosity of these latter systems has in general less pronounced
changes along the orbit with respect to the BeXBs, but can be subject to non-periodic variations on
time scales of seconds to hours, with $\Delta L_{\rm X} \sim 10-50$, 
due to hydrodynamic instabilities 
in the wind of the supergiant companion \citep{negueruela10}. 
A subclass of SGXBs, the supergiant fast X-ray
transients (SFXTs), spend a large fraction of their time \citep{romano11} in
a quiescent state (L$_X \sim $10$^{32}$-10$^{33}$~erg/s), and only sporadically undergo
bright outbursts ($\Delta$$L_{\rm X}$$\sim$10$^4$-10$^5$) lasting a  
few hours and reaching peak luminosities of $L_{\rm X}$$\sim$10$^{37}$~erg/s \citep{walter07}. 
The outbursts of SFXTs are associated to the accretion of particularly
dense clumps as in other SGXBs, but the origin of the lower persistent
luminosity and much more pronounced variability of these sources is
still a matter of debate \citep[see e.g.][and references therein]{zand05,bozzo11}. 

In this proceeding we report on the results of \xmm\ observations of four HMXBs
discovered with \inte\ and \beppo:
one classical SGXB (\jA), two candidates (\jD, \jE) and one confirmed (\jC) SFXTs.
We also report on the \xmm\
observation of the still unclassified \inte\ source \jB. 
A detailed description of the performed 
analysis can be found in \cite{bozzo12}.

\section{Results}
We used all the available \xmm\ observations of the five sources   
obtained through the guest observation programs awarded to our
research group.  
\xmm\ observation data files (ODFs) were processed
using the standard \xmm\ Science Analysis
System (v. 11.0)
Lightcurves were background-subtracted and barycenter-corrected;
spectra were background subtracted, rebinned, and where
required barycenter-corrected as well.

\subsection{\jA}
\jA\ is associated to the OB-V star SS~188, at a distance of
6.1~kpc \citep{maliziaatel}. Its X-ray spectrum
 could be fit with a simple power-law model ($N_{\rm
  H}$=1.5$\times$10$^{22}$~cm$^{-2}$, photon index $\Gamma$=2).
The X-ray luminosity of the source was
$\sim$2.4$\times$10$^{33}$~erg/s \citep{maliziaatel}.

\begin{figure}
\centering
\includegraphics[width=0.45\textwidth]{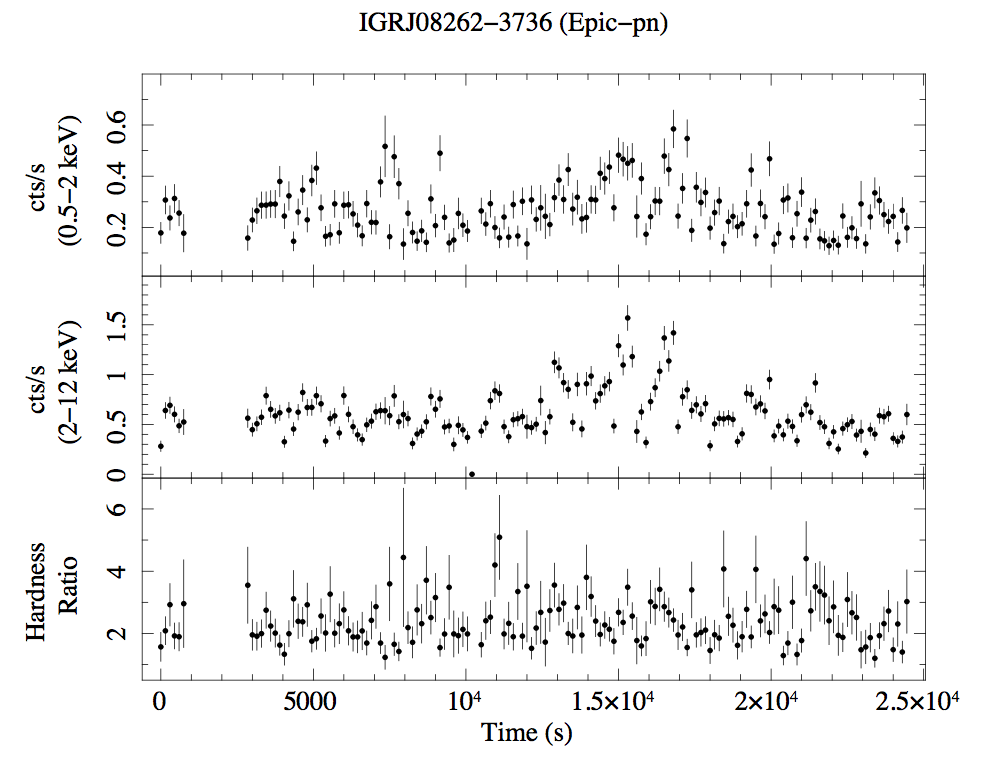}
\includegraphics[height=0.35\textwidth]{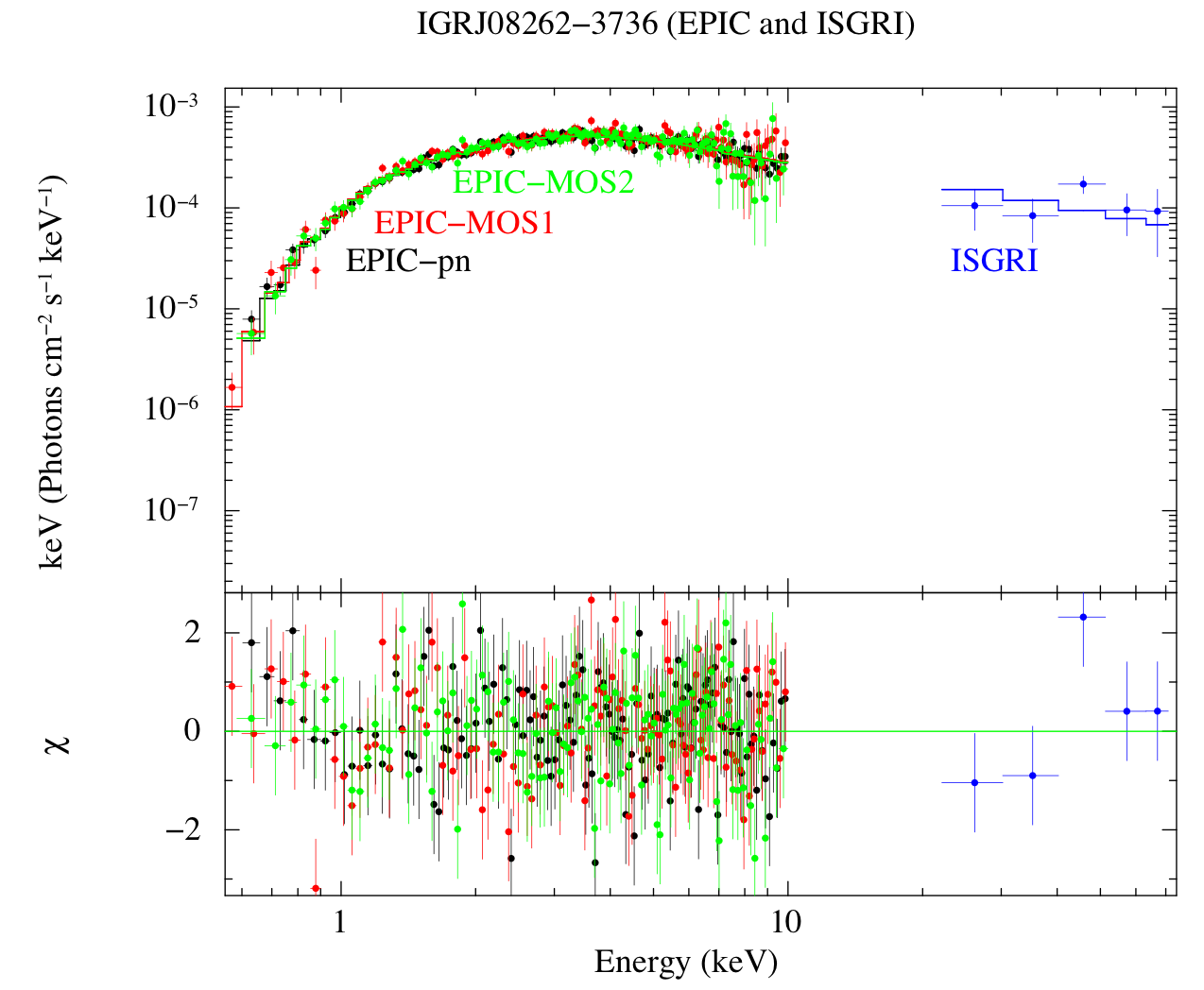}
\caption{ \footnotesize  {\it Left:} \xmm\ lightcurve of \jA\ in the 0.5-2~keV and 2-12~keV energy bands and the
   corresponding hardness ratio. {\it Right:} Average Epic-pn and \inte\,/ISGRI spectrum of \jA.\  
The best-fit model (solid line) is obtained with a partially covered power-law model. The residuals from the fit 
are shown in the bottom panel.}
\label{fig:jAlcurve} 
\end{figure}

We obtained an \xmm\ observation of \jA\ on 2010 October 16
for a total exposure time of $\sim$25~ks, with all three EPIC cameras 
operated in full frame mode. 
During this observation, the source displayed a moderate variability, with two relatively small
flares occurring about 1.5$\times$10$^4$~s  
after the beginning of the observation. 
The hardness ratio of the source 
(see Fig.~\ref{fig:jAlcurve}) does not show
clear variations in the source spectral properties  
with the count-rate. 
The Epic-pn spectrum of the source can not be fit by a simple
powerlaw model, and requires either an additional blackbody component at low 
energies, or the introduction of a 
partial covering (see Fig.~\ref{fig:jAlcurve}).
We also analyzed the mean spectrum obtained 
from the long-term monitoring of the source with ISGRI (obtained
through the HEAVENS online tool\footnote{http://www.isdc.unige.ch/heavens/}).
The XMM and INTEGRAL spectra can be fitted simultaneously, by
means of the same model described above, showing that the source 
has a virtually constant persistent X-ray flux. 
These timing and spectral 
behavior are typical of classical SGXBs.
The presence of ''soft excess'' in the spectrum is believed to be  
an ubiquitous feature of binary systems hosting accreting NSs
\citep{hickox04}.
In wind accreting binaries with X-ray luminosities comparable to that of \jA\ 
($\simeq$3$\times$10$^{34}$~erg/s, assuming a distance of 6.1~kpc),
the soft component  
is most likely originated by thermal X-ray photons close to the NS surface 
or by photoionized or collisionally heated diffuse gas in the binary
system.
As an alternative interpretation, 
also a partial-covering model would provide an
acceptable description of the X-ray spectrum. In this case, the soft excess would
be produced by the effect  
of partial obscuration of the emission from the NS by the surrounding
high-density material \citep[see e.g.,][]{tomsick09b}. 

\subsection{\jD}
\jD\ was discovered with \inte\ in 2006 \citep{kuulkers06}.
Long term monitoring of the source carried out with the \inte/ISGRI
and \swift/BAT showed that it is a weak persistent emitter in the hard
X-rays (average 18-60~keV flux of 
1.1~mCrab), displaying only sporadically 
flares with duration from few hours to $\sim$1~day. This behavior, together with a periodic modulation
detected in the hard X-ray data at nearly 8.45~days, led to the
interpretation of \jD\ as a HMXBs, possibly a SFXT \citep{dai11,sguera11}.
In the soft X-ray domain, \jD\ was observed with
\swift/XRT and \chan, showing two sources within 
its \inte\ error circle \citep[S1 and S2, ][]{vercellone09}.
A more pronounced variability of S1 (=CXOU~J173527.5-325554)
 led to the conclusion that this source is the most likely counterpart to 
\jD\ \citep[see e.g.][]{tomsick09}.

\xmm\ observed the region of \jD\ on 2011 March 6 for a
total effective exposure time of 20~ks.
MOS1 and pn cameras were operated in full-frame, while
MOS2 was in small-window mode. 
Among the two candidate soft X-ray counterparts of \jD,  in our XMM
observation only S2 was clearly detected, at a flux level of 
2.4$\times$10$^{-13}$~\ferg\ in the 0.5-10~keV range, compatible with
previous measurements \citep{vercellone09}.
No significant evidence for coherent modulations was found in the Epic
data, thus supporting the idea
that this is a persistent object  
unrelated with the \inte\ source. 
We could obtain a 3$\sigma$
upperlimit for S1 of $7\times10^{14}$ \ferg, a factor of $\sim$6  lower than
previous estimations. 
The increased dynamic range of S1/\jD\ gives strength to
the SFXT interpretation of the source.

\subsection{\jE}
\citet{corbet10} found an orbital period of $\sim$10~d from the hard
X-ray transient \inte\ source \jE,
hinting towards an interpretation as HMXB, possibly a supergiant system.  
\inte/ISGRI detected  two outbursts lasting a few
hours \citep{fiocchi10} from this source. On both occasions the ISGRI
spectrum could be fit with a simple 
power-law model ($\Gamma$$\simeq$2-2.6) with a flux of
2-3.3$\times$10$^{-10}$~erg~cm$^2$~s$^{-1}$ (20-50~keV). The source
was not detected at quiescent level, and the 3$\sigma$ upper limit
showed that \jE\ has a dynamical range of at least $100$.
The fast flaring behavior of the source, and its spectral properties,
suggested that \jE\ is a further member of the SFXT class 
discovered with \inte.

\begin{figure}
\centering
\includegraphics[height=0.35\textwidth]{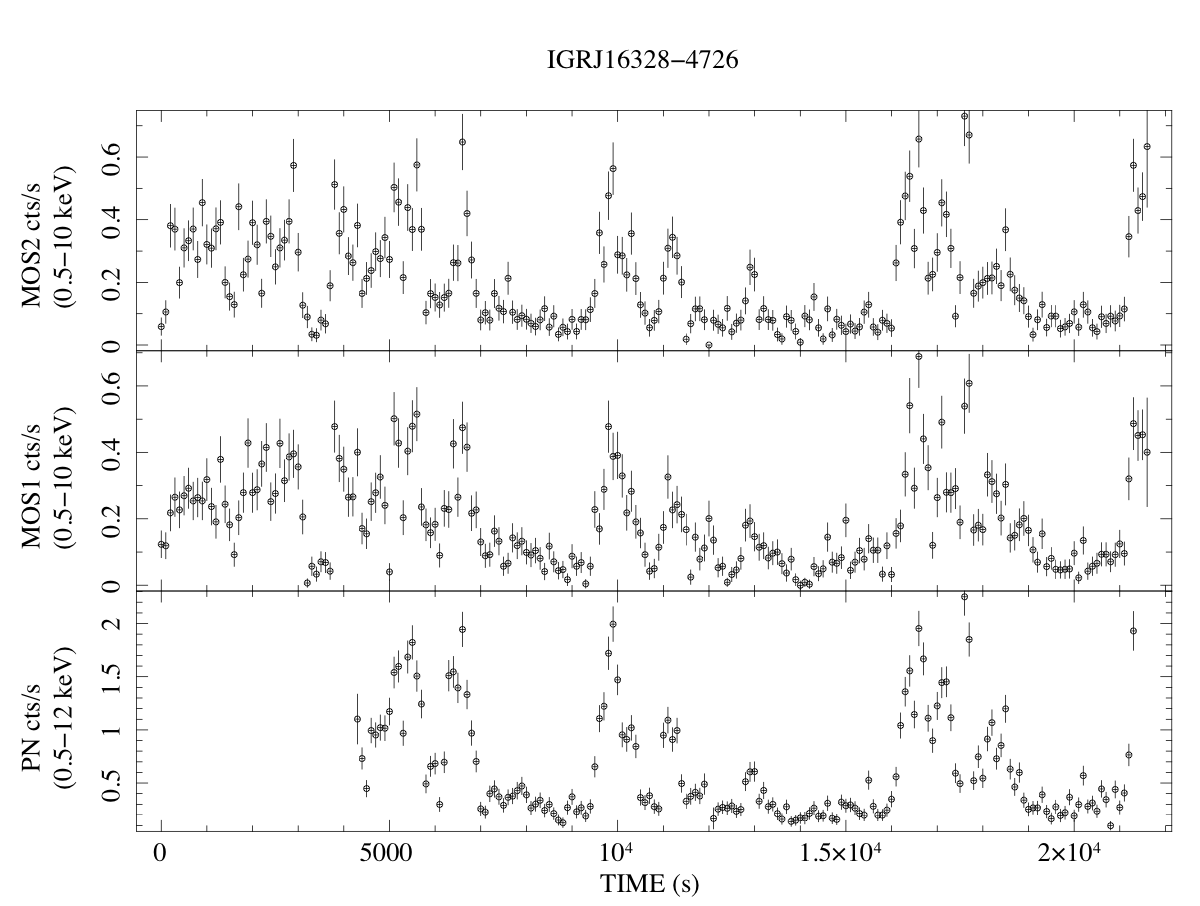}
\includegraphics[height=0.34\textwidth]{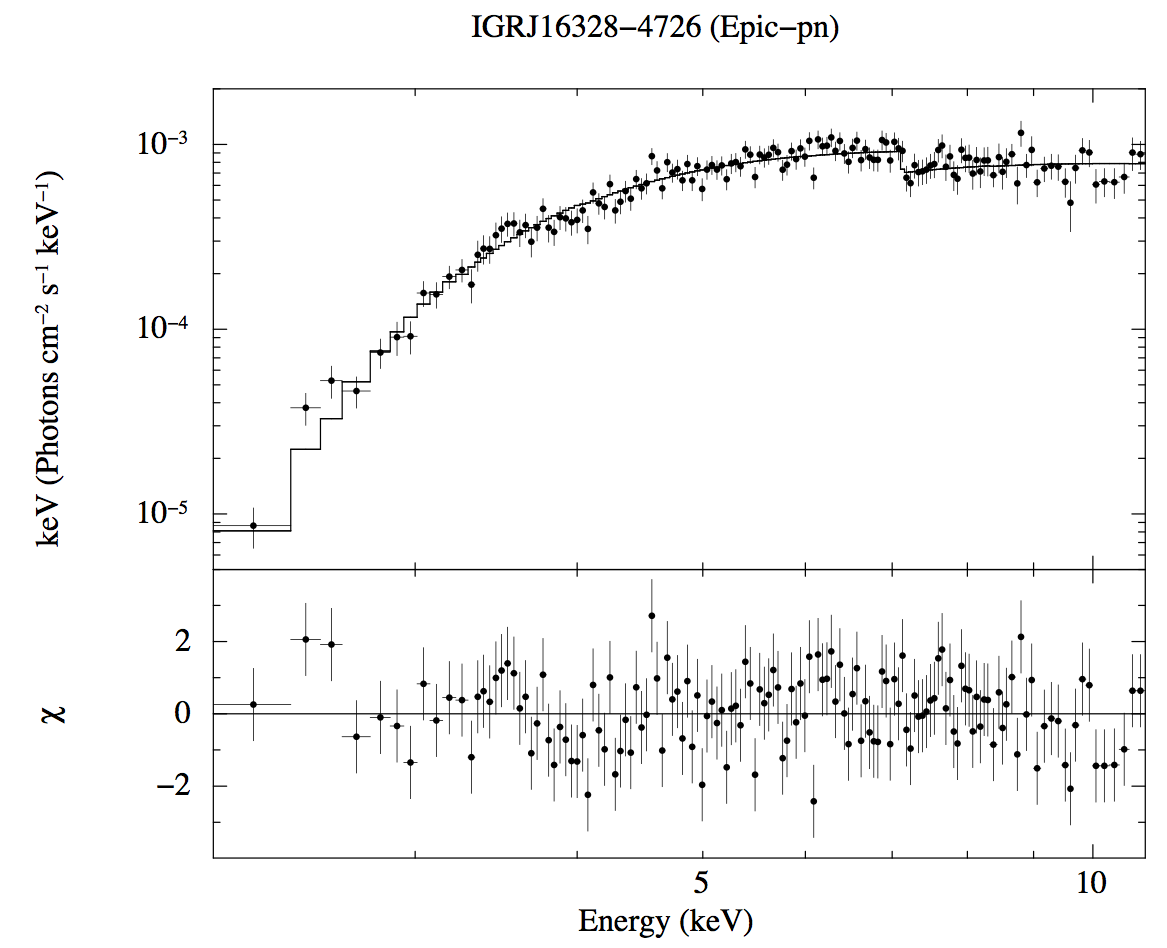}
\caption{\footnotesize  {\it Left:}\xmm\ 
  lightcurve of \jE\ in the 0.5-12~keV and 0.5-10~keV 
energy range for the Epic-pn and Epic-MOS, respectively. The time bin
is in all cases 100~s. {\it Right:} XMM/pn spectrum of \jE.  The residuals from a fit
with absorbed powerlaw model are shown in the bottom panel.}    
\label{fig:jElcurve} 
\end{figure}

\jE\ was observed by \xmm\ on 2011 February 20 for a total 
effective exposure time of 
14.7 (21)~ks for the Epic-pn (MOS) cameras.
The source displayed a clear
variability, with relatively small flares  
($\Delta$L$_{\rm X}$$\lesssim$10) occurring during periods of
low-level X-ray activity (Fig.~\ref{fig:jElcurve}). These flares closely resemble
the ones observed from the SFXT prototypes 
\citep{bozzo10}, and therefore are in further support to the association of \jE\ with this
class of objects.
The average Epic-pn spectrum extracted  
during the observation is well described by a strongly absorbed
power-law model ($N_{\rm  H} \sim 17.5 \times 10^{22} \textrm{cm}^{-2}$,
$\Gamma$=1.5$\pm$0.1; see Fig.~\ref{fig:jElcurve}). 
The flux ranges from $6.4\times10^{-12}$ (quiescent level) up to $5\times10^{-11}$ \ferg.
At odds with other SFXTs observed in quiescence, we do not
detect any soft excess in the X-ray spectrum of the source.  
This spectral component might have gone undetected in \jE\  
due to its relatively high absorption column density 
and the lower exposure time with respect to \xmm\ observations of
other quiescent SFXT sources with similar X-ray luminosity \citep{bozzo10}. 

\subsection{\jC} 
\jC\ is a confirmed SFXT source, discovered by
\beppo\ on 1998 during a 2-hours period of intense 
X-ray activity \citep{zand98}. Since then, several
outbursts from the source have been recorded with \inte\ and \swift\ \citep[see e.g.][]{sidoli09}. 
The source is at distance of 2.5~kpc, and has an orbital period of
30$\pm$0.1~d \citep{bird09}. Most of the observed outbursts took
place when the NS is close to the periastron 
\citep{zurita09}, however, outbursts in several periastron passages
were missing. 
This behavior might be explained by assuming that the NS in \jC\ has a 
highly eccentric orbit and is sporadically accreting  
mass from the clumpy wind of its supergiant companion.
The eccentricity of 0.3-0.4 inferred for \jC\ however 
can hardly support this scenario \citep[see e.g.][]{bird09}.

We obtained an \xmm\ observation of the region around \jC\ on 2010
March 21 for a total exposure time of 45~ks, at 
orbital phase 0.53$\pm$0.18 -- close to apastron \citep[with ephemeris given
by ][]{bird09}.
The Epic-pn camera was operated in
Full Frame mode, while the MOS1 
and MOS2 were operated in Small Window and Fast Uncompressed
mode, respectively.  
\jC\ was not detected with the Epic cameras.
We estimated from these data a 3-$\sigma$ upper 
limit on the source luminosity of $\sim$2$\times$10$^{32}$~erg/s, 
comparable with the one estimated
by the previous non-detection of the source at phase 0.51 \citep{bozzo08b}. 
The mechanism involved in the prolonged quiescent state of \jC\
 is not known yet. Proposed models include inhibition of
accretion by the NS, due to centrifugal and/or magnetic barrier \citep{bozzo08b};
or alternatively accretion from a highly rarefied and structured
companion wind \citep[see, e.g.][and references therein]{zurita09}.

\subsection{\jB} 
\jB\ is an unclassified \inte\ source \cite{bird09}.
The estimated fluxes in the 20-40~keV and 40-100~keV energy bands are
$0.3\pm0.1$ mCrab and $0.9\pm0.1$~mCrab, respectively.

\begin{figure}
\centering
\includegraphics[width=0.45\textwidth]{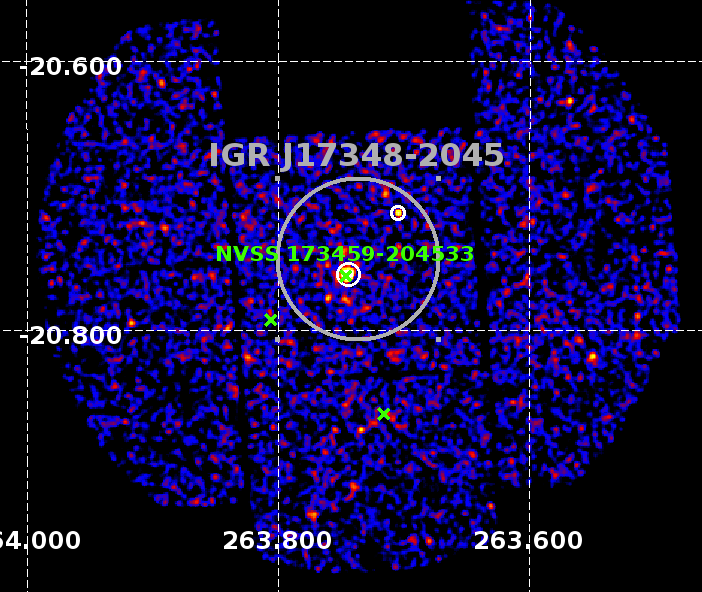}
\caption{ \footnotesize Epic/MOS image of \jB\ in the
  0.5-10~keV band. The two soft X-ray counterparts of \jB\  
detected are indicated as white circles. The radio object NVSS\,J173459-204533, spatially
coincident with XMMU\,J173458.8-204530, is also indicated in green.}    
\label{fig:jBima} 
\end{figure}

The source was observed by \xmm\ on 2011 March 2 for a total effective
exposure time of 10 (4) ksec for Epic-MOS (pn). 
The three EPIC detectors were operated in Full Frame mode. 
Inside the \inte\ error circle we detect two soft X-ray
sources (named XMMU\,J173458.8-204530 and
XMMU\,J173449.2-204244, according to the \xmm\ convention).
The former source is brighter and located close to the center of the \inte\
error-circle; the latter
is too dim to extract any meaningful spectral information
(see Fig.~\ref{fig:jBima}).  We assume in the following  
that XMMU\,J173458.8-204530 is the true counterpart to \jB.
Its XMM spectrum is modeled as a strongly absorbed PL 
(N$_H$ $\sim 10^{23}$ cm$^{-2}$; $\Gamma = 1.5$).
A simultaneous fit with the long-term ISGRI spectrum (obtained from
HEAVENS) is compatible with the source being a persistent hard X-ray emitter. 
The uncertainty in the Epic-MOS position does not allow a firm
identification of counterparts at other wavelengths,  
but we remark here the possible association with the radio source
NVSS\,J173459-204533. If this association will be confirmed by future  
observations, we argue that \jB\ could be one of the highly absorbed active  
Galactic nuclei (AGN) discovered with \inte\ \citep[see e.g. ][and references therein]{ricci11}.

\section{Conclusions}
We analyzed in the 1-10~keV region five hard X-ray sources, through
dedicated XMM-Newton observations \citep[published in][]{bozzo12}. 
The observations provided the first detailed spectral and timing study
of IGR J08262-3736, showing signatures of an accreting neutron star. 
The variability of the 1-10~keV counterpart to IGR J17354-3255,
assessed through the deepest upperlimit of the source, favors its
interpretation as SFXT. A similar interpretation is strengthened for IGR
J16328-4726 thanks to the observed time variability. 
The mechanism involved in the prolonged quiescent state of SAX
J1818.6-1703, and SFXTs in general, is still matter of debate.  Thanks
to the first 1-10~keV study of IGR J17348-2045 we could identify its
soft X-ray emission, and localize a possible radio counterpart. We
suggest for this source an extragalactic origin. 

\begin{multicols}{2}
\footnotesize{

}
\end{multicols}


\begin{thebibliography}{27}
\expandafter\ifx\csname natexlab\endcsname\relax\def\natexlab#1{#1}\fi

\bibitem[{{Bird} {et~al.}(2009){Bird}, {Bazzano}, {Hill}, {McBride}, {Sguera},
  {Shaw}, \& {Watkins}}]{bird09}
{Bird}, A.~J., {et~al.} 2009, MNRAS, 393, L11

\bibitem[{{Bozzo} {et~al.}(2008){Bozzo}, {Falanga}, \& {Stella}}]{bozzo08b}
{Bozzo}, E., {et~al.} 2008, ApJ, 683, 1031

\bibitem[{{Bozzo} {et~al.}(2011){Bozzo}, {Giunta}, {Cusumano}, {Ferrigno},
  {Walter}, {Campana}, {Falanga}, {Israel}, \& {Stella}}]{bozzo11}
{Bozzo}, E., {et~al.} 2011, A\&A, 531, A130

\bibitem[{{Bozzo} {et~al.}(2012){Bozzo}, {Pavan}, {Ferrigno}, {Falanga},
  {Campana}, {Paltani}, {Stella}, \& {Walter}}]{bozzo12}
{Bozzo}, E., {et~al.} 2012, A\&A, 544, 118

\bibitem[{{Bozzo} {et~al.}(2010){Bozzo}, {Stella}, {Ferrigno}, {Giunta},
  {Falanga}, {Campana}, {Israel}, \& {Leyder}}]{bozzo10}
{Bozzo}, E., {et~al.} 2010, A\&A, 519, A6

\bibitem[{{Corbet} {et~al.}(2010){Corbet}, {Barthelmy}, {Baumgartner}, {Krimm},
  {Markwardt}, {Skinner}, \& {Tueller}}]{corbet10}
{Corbet}, R.~H.~D., {et~al.} 2010,
  Astr. Tel., 2588

\bibitem[{{D'A{\`\i}} {et~al.}(2011){D'A{\`\i}}, {La Parola}, {Cusumano},
  {Segreto}, {Romano}, {Vercellone}, \& {Robba}}]{dai11}
{D'A{\`\i}}, A., {et~al.} 2011, A\&A, 529, A30

\bibitem[{{Fiocchi} {et~al.}(2010){Fiocchi}, {Sguera}, {Bazzano}, {Bassani},
  {Bird}, {Natalucci}, \& {Ubertini}}]{fiocchi10}
{Fiocchi}, M., {et~al.} 2010, ApJl, 725, L68

\bibitem[{{Hickox} {et~al.}(2004){Hickox}, {Narayan}, \& {Kallman}}]{hickox04}
{Hickox}, R.~C., {et~al.} 2004, ApJ, 614, 881

\bibitem[{{in 't Zand} {et~al.}(1998){in 't Zand}, {Heise}, {Smith}, {Muller},
  {Ubertini}, \& {Bazzano}}]{zand98}
{in 't Zand}, J., {et~al.} 1998, IAUcirc, 6840, 2

\bibitem[{{in 't Zand}(2005)}]{zand05}
{in 't Zand}, J. 2005, A\&A, 441, L1

\bibitem[{{Kreykenbohm} {et~al.}(2008){Kreykenbohm}, {Wilms}, {Kretschmar},
  {Torrej{\'o}n}, {Pottschmidt}, {Hanke}, {Santangelo}, {Ferrigno}, \&
  {Staubert}}]{kreykenbohm11}
{Kreykenbohm}, {et~al.} 2008, A\&A, 492, 511

\bibitem[{{Kuulkers} {et~al.}(2006){Kuulkers}, {Shaw}, {Paizis}, {Gros},
  {Chenevez}, {Sanchez-Fernandez}, {Brandt}, {Courvoisier}, {Garau}, {Ebisawa},
  {Kretschmar}, {Markwardt}, {Mowlavi}, {Oosterbroek}, {Orr}, {Oneca}, \&
  {Wijnands}}]{kuulkers06}
{Kuulkers}, E., {et~al.} 2006, Astr. Tel., 874

\bibitem[{{Malizia} {et~al.}(2011){Malizia}, {Landi}, {Bassani}, {Bird},
  {Gehrels}, \& {Kennea}}]{maliziaatel}
{Malizia}, A., {et~al.} 2011, Astr. Tel., 3294

\bibitem[{{Negueruela}(2010)}]{negueruela10}
{Negueruela}, I. 2010, ASP Conf. Series, 422, 57 

\bibitem[{{Ricci} {et~al.}(2011){Ricci}, {Walter}, {Courvoisier}, \&
  {Paltani}}]{ricci11}
{Ricci}, C., {et~al.} 2011, A\&A, 532, A102

\bibitem[{{Romano} {et~al.}(2011){Romano}, {La Parola}, {Vercellone},
  {Cusumano}, {Sidoli}, {Krimm}, {Pagani}, {Esposito}, {Hoversten}, {Kennea},
  {Page}, {Burrows}, \& {Gehrels}}]{romano11}
{Romano}, P., {et~al.} 2011, MNRAS, 410,
  1825

\bibitem[{{Sguera} {et~al.}(2006){Sguera}, {Bazzano}, {Bird}, {Dean},
  {Ubertini}, {Barlow}, {Bassani}, {Clark}, {Hill}, {Malizia}, {Molina}, \&
  {Stephen}}]{sguera06}
{Sguera}, V., {et~al.} 2006, ApJ, 646, 452

\bibitem[{{Sguera} {et~al.}(2011){Sguera}, {Drave}, {Bird}, {Bazzano}, {Landi},
  \& {Ubertini}}]{sguera11}
{Sguera}, V., {et~al.} 2011, MNRAS, 417, 573

\bibitem[{{Sidoli} {et~al.}(2009){Sidoli}, {Romano}, {Esposito}, {Parola},
  {Kennea}, {Krimm}, {Chester}, {Bazzano}, {Burrows}, \& {Gehrels}}]{sidoli09}
{Sidoli}, L., {et~al.} 2009, MNRAS, 400, 258

\bibitem[{{Stella} {et~al.}(1986){Stella}, {White}, \& {Rosner}}]{stella86}
{Stella}, L., {et~al.} 1986, ApJ, 308, 669

\bibitem[{{Tomsick} {et~al.}(2009{\natexlab{a}}){Tomsick}, {Chaty},
  {Rodriguez}, {Walter}, \& {Kaaret}}]{tomsick09}
{Tomsick}, J.~A., {et~al.} ApJ, 701, 811

\bibitem[{{Tomsick} {et~al.}(2009{\natexlab{b}}){Tomsick}, {Chaty},
  {Rodriguez}, {Walter}, {Kaaret}, \& {Tovmassian}}]{tomsick09b}
{Tomsick}, J.~A., {et~al.} 2009{\natexlab{b}},
  ApJ, 694, 344

\bibitem[{{Torrej{\'o}n} {et~al.}(2010){Torrej{\'o}n}, {Negueruela}, {Smith},
  \& {Harrison}}]{torrejon10}
{Torrej{\'o}n}, J.~M., {et~al.}  2010, A\&A, 510, A61

\bibitem[{{Vercellone} {et~al.}(2009){Vercellone}, {D'Ammando}, {Striani},
  {Tavani}, {Sabatini}, {Bulgarelli}, {Gianotti}, {Trifoglio}, {Feroci},
  {Lazzarotto}, {Del Monte}, {Pittori}, {Verrecchia}, {Pellizzoni}, {Pilia},
  {Chen}, {Giuliani}, {Piano}, {Pucella}, {Vittorini}, {Costa}, {Donnarumma},
  {Pacciani}, {Soffitta}, {Evangelista}, {Lapshov}, {Rapisarda}, {Argan},
  {Trois}, {de Paris}, {Marisaldi}, {Di Cocco}, {Labanti}, {Fuschino}, {Galli},
  {Caraveo}, {Mereghetti}, {Perotti}, {Fiorini}, {Zambra}, {Barbiellini},
  {Longo}, {Moretti}, {Vallazza}, {Picozza}, {Morselli}, {Prest}, {Lipari},
  {Zanello}, {Cattaneo}, {Rappoldi}, {Santolamazza}, {Colafrancesco}, {Giommi},
  {Salotti}, {Romano}, {Burrows}, \& {Gehrels}}]{vercellone09}
{Vercellone}, S., {et~al.} 2009, 
  Astr. Tel., 2019

\bibitem[{{Walter} \& {Zurita Heras}(2007)}]{walter07}
{Walter}, R. \& {Zurita Heras}, J. 2007, A\&A, 476, 335

\bibitem[{{Zurita Heras} \& {Chaty}(2009)}]{zurita09}
{Zurita Heras}, J.~A. \& {Chaty}, S. 2009, A\&A, 493, L1

\end{thebibliography}
\end{document}